# The Cost of Problem-Based Learning: An Example in Information Systems Engineering

Vincent Ribaud, Philippe Saliou
*Département d'informatique ; Université de Brest; France*
*{ ribaud, psaliou }@univ-brest.fr*

*Abstract*

*High-quality education helps in finding a job - but student skills heterogeneity and student reluctance to move towards a professional attitude are important barriers to employability. We re-engineered some of the technical courses of a Masters in software development using a Problem-Based Learning (PBL) approach. Although initial results are encouraging, the cost of using PBL must be taken into account. Two aspects are particularly expensive: (i) set-up of the software development practicum, a mid-sized information system and its environment; (ii) screenwriting of problem-based learning scenarios, including procurement of input artefacts.*

## 1. Introduction

The information technology industry in Morocco is growing fast, thanks to offshore practices. Moroccan government funding for IT education helped start new programmes entitled "Masters Offshoring" at almost every Moroccan university. Based on a previous cooperation, the University of Brest established in 2010 a joint Masters degree with nine Moroccan state universities, focused on software development. We introduced a Problem-Based Learning (PBL) approach within some of the programme courses, mainly in an attempt to resolve two problems: heterogeneity of knowledge and skills among students, and a reluctance on the part of certain students to move from a passive learning attitude to one that is active. Several meta-analyses conducted beyond medical education [1, 2] report positive effects from PBL on the skills of students; [3] lists several applications of PBL in professional sciences reported as being time-consuming - but we are not aware of any attempt to highlight the financial burden of a PBL approach. This paper aims to describe the cost of PBL, which may turn out to be much higher than expected. General issues are discussed in section 2. Section 3 presents the practicum in which the PBL is run, and section 4 some considerations about the effort involved in running the PBL sessions. We finish with a conclusion.

## 2. Background

### 2.1. Origin of students

The first year of Masters is performed in Morocco in four quite different specialties (Software Development and Quality, Networking and Systems, Information System Engineering and Applied Informatics Offshoring) at 9 universities (Agadir, Casablanca, Fez, Kenitra, El Jadida, Marrakech, Rabat, Settat and Tanger). The second year of study takes place in France: 6 months of study in Brest, followed by a period of 6 months in France, with pre-employment in Morocco. The 2010-2011 cohort comprised 31 students, the 2011-2012 cohort 35 students, and the current cohort comprises 27 students. Despite the fact that the selection level is set as high as possible, knowledge and skills acquired by the end of the first year can vary from student to student, raising a heterogeneity problem – this was the first reason behind out trying out the PBL approach reported here.

### 2.2. Content of the second year

The programme curriculum has been designed to train engineers in the development (design, production and maintenance) of software projects. During the first semester of the second year of the Masters (September to March) all students attend 8 technical courses: database and Java programming, development environments, object-oriented design, distributed systems, web technologies, software engineering, information systems, and J2EE development. They also attend courses in English and Communication in French, and a general introduction to offshore context. The 6-month internship takes place from April to September. For almost all Moroccan students, the internship in France is their first encounter with the industrial world and its expectations. Some interns experience difficulty in adopting a professional attitude and in leaving their student wardrobe at home. Preparing students for the real world was the second reason behind the introduction of the PBL experience.

### 2.3. The need for a different approach in information systems engineering

Prince [4] defines Problem-Based Learning as "*an instructional method where relevant problems are introduced at the beginning of the instruction cycle and used to provide the context and motivation for the learning that follows. It is always active and usually (but not necessarily) collaborative [...]. PBL typically involves significant amounts of self-directed learning on the part of the students.*" In 2010 and 2011, we tried a range of approaches aimed at resolving the problem of heterogeneity between students, and came to the conclusion that students have to identify what they are lacking, and fill the gap mainly by themselves - a good reason to use PBL as a learning vehicle. Also, PBL is an active learning approach intended to foster problem-solving skills and autonomy - the second issue we were faced with.

### 2.4. Setting of the PBL approach

We selected Information Systems courses to be taught using PBL (ideally) regardless of who is in charge of the course. The selected courses are: Database and Java Programming (48h), Software Engineering (60h), and Information Systems (60h): a total of 168 hours – one third of the technical courses as a whole. These are taught by three professors, including the author of this paper. The initial learning period (2 months) is taught using the PBL approach exclusively, in hope of bootstrapping the rest of the year. We did make several design choices. PBL group size is limited to 14; because of this, we have to duplicate course hours. Insofar as possible, students work alone. The same background - a medium-sized IS - is used throughout all sequences of all courses involved. Because PBL favours an inductive approach, we present the development cycle in an ascendant fashion – from code to requirements analysis. With a few pragmatic exceptions, these are general design choices. The faculty head does not allow us to duplicate all of the hours; 60 of the 168 hours must be taught to the whole class of 27 students at once. Students may work in pairs, essentially for economic reasons. Learning-by-doing tutorials may well make use of other, smaller cases, but the knowledge gained will be reused and re-contextualized into the main case.

## 3. The practicum

Project-based learning is frequently used in engineering education. Perrenet et al. [5] state that "*PBL offers good prospects in the first few years of a programme, but in later phases project work offers a strong alternative.*" The differences that they noted [5] included (i) project tasks are closer to professional reality and therefore take a longer period of time than PBL problems; (ii) project work is more directed to the *application* of knowledge, whereas problem-based learning is more directed to the *acquisition* of knowledge; (iii) management of time and resources by the students as well as task and role differentiation is very important in project-based learning. Blending project-based and problem-based learning could turn out to be fruitful: a compromise between learning and doing. We took the opportunity to use an existing IS as the practicum where any PBL activities are run – anchoring activities in reality and reusing existing artefacts as required - to alleviate time pressure and resource management. Information System (IS) is the main subject of the PBL approach, and a practical understanding of the development cycle of an IS is an underlying objective of any PBL session. It will gradually be revealed to students that each PBL session is contributing to some extent to the development of a new sub-system. But unlike project-based learning, neither the new sub-system nor its development cycle are the target of the PBL sessions. Its role is merely a contribution to setting up a different context for each session and yielding input artefacts and learning objectives.

### 3.1. Architecture

A Management Information System, called SIGILI, has been developed to meet the needs of our Informatics Department. SIGILI was designed to manage schooling and was used by administrative staff and programme managers. SIGILI is composed of 3 sub-systems: a schooling management system; an internships management system; and a competencies management system. The three sub-systems use a three-tier architecture based on Java/JSP, Tomcat or Oracle AS as application servers and Oracle as the database management system. The whole system was developed from 2005 to 2007 with the author acting as project manager (the job he had performed for 13 years prior to joining our university); each sub-system was developed by a team of 5-6 full-time interns during their 7-month Masters internship (17 interns overall). The system and its technical environment will be used throughout all PBL sequences as the practicum in which activities take place.

### 3.2. Legacy, complexity and heterogeneity

A major challenge for IT students is dealing with the complexity and heterogeneity of legacy systems. Information systems are built through successive projects, with people, processes and technologies changing over time. "*Problem-based learning can help students to learn with complexity, to see that there are no straightforward answers to problem scenarios, but that learning and life takes place in contexts, contexts which affect the kinds of solutions that are available and possible* [6]." The SIGILI system has a real-scale that provides us with all the required situations – straightforward to complex – we need to set up PBL sessions.

### 3.3. Example of sessions

As mentioned in the design principles, we inverted the development cycle so that during the first period, PBL sessions are related to delivery and programming activities. Examples of PBL sessions for the Database Server are:
- Restructure a set of Data Definition Language (DDL) scripts in a design-based hierarchy
- Check consistency between code artefacts and technical detailed specification
- Refactor a complete sub-system according to naming and organization rules

Examples of PBL sessions for the Application Server are:
- Retro-design then develop Web pages with the user's manual yielded as specifications
- Perform a code review of an existing module

At this point, PBL is related to architecture and detailed design.

Examples of sessions are:
 • Retro-engineering of a relational model into an entity-association model, then correcting, enhancing and completing it so as to be able to generate the relational model again
 • Establishing a high-level view of services provided by software components, and providing an architectural hierarchy of the system, its sub-systems and components

### 3.4. Assessment

As is true of most all active learning approaches, PBL assessment is part of the PBL itself - this is what J. Tardif [7] calls "*the entrenchment of assessment in learning*". When a PBL session artefact is delivered, the tutor examines it and provides feedback about certain points to be improved upon or started over. But the workload may be too heavy and we also practice a stop-and-go approach: it works or it does not work. In the latter case, students are awarded a low mark - but are provided with a working artefact that allows them to continue their work.

## 4. The cost of PBL

### 4.1. A mid-sized Information System

Recall that SIGILI development was performed by 17 interns over a period of three years. Since intern productivity is roughly estimated to be half that of a novice engineer, the development workload is estimated at 17*7 / 2 = 60 man/months. The project management workload is estimated at 12 man/months over the 3-year period. That is a 6 man/year effort, and since we mainly use one of the three sub-systems, we can estimate that a 2 man/year workload is required to provide the PBL approach with a real-case setting.

### 4.2. SIGILI artefacts

Successive phases of SIGILI development produced an exhaustive set of major deliverables issued in a software project at our disposal. These project artefacts were obviously not originally designed to serve the PBL approach, which we built only recently, and most have to be reworked before they can be used in a PBL setting. PBL sessions are played within software development phases such as design, coding or tests where output artefacts of one phase are used as input artefacts for the next. Successive cases should rely on sound and complete artefacts, even though they should, ideally, have been produced by students. There is no choice: scenario designers have to produce good artefacts themselves to accompany the case; otherwise tutors will find themselves unable to run the case. And it is a highly time-consuming task. We roughly estimate that a full working week is required to produce the input artefacts used in a 4-hour PBL session.

### 4.3. Problem design

"*In problem-based curricula the problem scenarios should serve as the central component of each module* [8]." Curricular content has to be organized around problem scenarios rather than subjects or disciplines. The complexity of problem design is a challenge to many tutors implementing problem-based learning. As mentioned by [8]: "*Are we asking them to solve a closed problem by using linear problem-solving techniques? Or are we asking them to do something very different, such as using their experiential and propositional knowledge to manage the problem situation?* " Both design approaches are expensive - the former because we have to firmly drive students towards the expected solution, the latter because we have to try to envisage a whole set of solutions that might be chosen by students and their consequences for problem-solving and student learning. Based on their experience, teachers agree that designing and preparing a 4-hour PBL session requires 1 week of full-time work.

### 4.4. The entry ticket to PBL in information system engineering

Based on the estimations given above, we are able to estimate the cost of our PBL approach. A 4-hour PBL session requires 1 week for design and screenwriting; and 1 week for production of the input artefacts – meaning that one PBL hour demands a 2.5 day effort. Each teacher will spend a different amount of time preparing new lectures or new labs, but a full working day to prepare 1 lecture hour, and 1.5 days for one lab hour seems an accurate estimation. Hence, the initial set-up of a PBL approach is around twice as costly as a traditional approach - and this does not include the development cost of the SIGILI system used in all PBL sessions. In [9], authors compare teaching time in medical education between a traditional teacher-centred, subject-oriented curriculum and a problem-based, student-centred curriculum. They conclude that "*there were no differences in the total amount of teaching time required by each of the two curricular approaches to medical education. There were, however, major differences in how teachers spent their teaching time.* [9]"

## 5. Conclusion

In a joint Masters degree, student heterogeneity and lack of industrial experience presented us with new challenges. A PBL approach was trialled on a few courses in order to develop a reflective practice. Although PBL has proved its worth in engineering education [10], our experience is too limited to draw any conclusions about student perception of PBL or the pros and cons of the approach. However a comparison between the cost of a PBL approach and a traditional teaching based on lectures and labs tends to show that PBL is twice as costly as the usual approach. This is not correlated with a study in medical education, but the PBL curriculum focuses upon small-group tutorial learning which probably multiplies the number of teaching hours compared to traditional medical education. This paper therefore argues that the price of starting a PBL approach in information system engineering is high - an investment faculty heads or colleagues might not want to make.